\documentclass{article}

\usepackage{graphicx}
\usepackage{amsmath}

\usepackage{hyperref} 
\usepackage[linewidth=1pt]{mdframed}
\usepackage{booktabs}
\usepackage{float}
\usepackage{natbib}
\def\bSig\mathbf{\Sigma}

\DeclareSymbolFont{extraup}{U}{zavm}{m}{n}
\DeclareMathSymbol{\varheart}{\mathalpha}{extraup}{86}
\DeclareMathSymbol{\vardiamond}{\mathalpha}{extraup}{87}

\begin{document}

\title{}

\vspace*{\stretch{1.0}}
   \begin{center}
      \Large\textbf{A reckless guide to P-values}\\
      \bigskip
      \Large{Local evidence, global errors}\\
      \bigskip
     \small{Chapter 13 in the book Good Research Practice in Experimental Pharmacology, editors A. Bespalov, MC Michel, and T Steckler, to be published by Springer.
     
     Open Access, Creative Commons 4.0}
   \end{center}
   \vspace*{\stretch{2.0}}

\author{\textbf{Michael J. Lew}\\
Department of Pharmacology and Therapeutics\\
University of Melbourne}

\maketitle

\section{Abstract}

This chapter demystifies P-values, hypothesis tests and significance tests, and introduces the concepts of local evidence and global error rates. The local evidence is embodied in \textit{this} data and concerns the hypotheses of interest for \textit{this} experiment, whereas the global error rate is a property of the statistical analysis and sampling procedure. It is shown using simple examples that local evidence and global error rates can be, and should be, considered together when making inferences. Power analysis for experimental design for hypothesis testing are explained, along with the more locally focussed expected P-values. Issues relating to multiple testing, HARKing, and P-hacking are explained, and it is shown that, in many situation, their effects on local evidence and global error rates are in conflict, a conflict that can always be overcome by a fresh dataset from replication of key experiments. Statistics is complicated, and so is science. There is no singular right way to do either, and universally acceptable compromises may not exist. Statistics offers a wide array of tools for assisting with scientific inference by calibrating uncertainty, but statistical inference is not a substitute for scientific inference. P-values are useful indices of evidence and deserve their place in the statistical toolbox of basic pharmacologists.

\tableofcontents

\section{Introduction}

There is a widespread consensus that we are in the midst of a `reproducibility crisis' and that inappropriate application of statistical methods facilitates, or even causes, irreproducibility \citep{Ioannidis:2005bw,NuzzoPvals,Colquhoun:2014cv,George:2017fn,Wagenmakers:2018iz}. P-values are a ``pervasive problem'' \citep{Wagenmakers:2007ur} because they are misunderstood, misapplied, and answer a question that no-one asks \citep{RoyallBook,HalseyFickleP,Colquhoun:2014cv}. They exaggerate evidence \citep{Johnson:2013eh,Benjamin:2018gh} or they are irreconcilable with evidence \citep{Berger:1987tf}. What's worse, `P-hacking' amplifies their intrinsic shortcomings \citep{FraserQRP:2018}. The inescapable conclusion, it would seem, is that P-values should be eliminated by replacement with Bayes factors \citep{Goodman:2001tx,Wagenmakers:2007ur} or confidence intervals \citep{Cumming:2008wb}, or by simply doing without \citep{doi:10.1080/01973533.2015.1012991}. However, much of the blame for irreproducibility that is apportioned to P-values is based on pervasive and pernicious misunderstandings.

This chapter is an attempt to resolve those misunderstandings. Some might say it is a reckless attempt because history suggests that it is doomed to failure, and reckless also because it goes against much of the conventional wisdom regarding P-values and will therefore be seen by some as promoting inappropriate statistical practices. That's OK though, because the conventional wisdom regarding P-values is mistaken in important ways, and those mistakes fuel false suppositions regarding what practices are appropriate.

\subsection{On the role of statistics}

Statistics is complicated\footnote{Even its grammatical form is complicated: ``statistics'' looks like a plural noun, but it both plural when  referring to values calculated from data and singular when referring to the discipline or approaches to data analysis.} but is usually presented simplistically in the statistics textbooks and courses studied by pharmacologists. Readers of those books and graduates of those courses should therefore be forgiven for wrongly assuming that statistics is a set of rules and recipes that must be applied in order to obtain a statistically valid statistically significant. The instructions say that you match the data to the recipe, turn the crank, and bingo: it's significant, or not. If you do it right then you might rewarded with a star! No matter how explicable that simplistic view of statistics might be, it is far too limiting. It leads to thoughtless use of a limited set of methods and to over-reliance on the familiar but misunderstood P-value. It prevents the full utilisation of statistical thinking within scientific inference, and allows bad statistics to license false inferences. We have to aim for more than the rote-learning of recipes in statistics courses because while statistics is not simple, good science is harder. I therefore take as a working assumption the notion that good scientists are capable of dealing with the intricacies of statistical thinking.

I will admit up front that it is not essential to have a statistical inference in order to make a scientific inference. For example, there is little need for a formal statistical analysis if results can be dealt with using the inter-ocular impact test\footnote{In other words, results that hit you right between the eyes. In the Australian vernacular the inter-ocular impact test is the bloody obvious test.}. However, scientific inferences can be made more securely with statistics because it offers a rich set of tools for calibrating uncertainty. Statistical analysis is particularly helpful in the penumbral `maybe zone' where the uncertainty is relatively evenly balanced---the zone where scientists are most likely to be swayed by biasses into over-interpretation of random deviations within the noise. The extra insight from a well-implemented statistical analysis can protect from the desire to find something notable, and thereby reduce the number of false claims made.

\begin{quote}
Most people need all the help they can get to prevent them making fools of themselves by claiming that their favourite theory is substantiated by observations that do nothing of the sort. 

\hfill ---\citep[p. 1]{ColquhounBook}	
\end{quote}

Improved utilisation of statistical approaches would indeed help to minimise the number of times that pharmacologists make fools of themselves by reducing the number of false positive results in pharmacological journals and, consequently, reduce the number of faulty leads that fail to translate into a therapeutic \citep{Begley:2012ga}. However, even ideal application of the most appropriate statistical methods would not improve the replicability of published results quite as much as might be assumed because not every result that fails to be replicated is a false positive and not every mistaken conclusion would be prevented by better statistical inferences.

Basic pharmacological studies are typically performed using biological models such as cell lines, tissue samples, or laboratory animals and so even if the original results are not false positives a replication might fail when it is conducted using different models,  \citep{Drucker:2016du}. Replications might also fail when the original results are critically dependent on unrecognised methodological details, or on reagents such as antibodies that have properties that can vary over time or between sources \citep{Berglund:2008kn,Baker:uq,Voelkl:2018gp}. It is those types of irreproducibility rather than false positives that are responsible for many failures of published leads to translate into clinical targets or therapeutics (see also Chapter 11). The distinction being made here is between false positive inferences which lack `internal validity' and failures of generalisability which lack `external validity' even though correct in themselves. It is an important distinction because the former can be reduced by more appropriate use of statistical methods but the latter can not.

The inherent objectivity of statistics can minimise the number of times that we make fools of ourselves, but just \textit{doing statistics} is not enough, because  it is not a set of rules for scientists to follow to make automated scientific inferences. To get from calibrated statistical inferences to reliable inferences about the real world, the statistical analyses have to be interpreted; thoughtfully and in the full knowledge of the properties of the tool and the nature of the real world system being probed. Some researchers might be disconcerted by the fact that statistics cannot provide such certainty, because they just want to be told whether their latest result is ``real''. No matter how attractive it might be to fob off onto statistics the responsibility for inferences, the answers that scientists seek cannot be answered by statistics alone. 

\section{All about P-values}

P-values are not everything, and they are certainly not nothing. There are many, many useful procedures and tools in statistics that do not involve or provide P-values, but P-values are by far the most widely used inferential statistic in basic pharmacological research papers.0

\begin{quote}
P-values are a practical success but a critical failure. Scientists the world over use them, but scarcely a statistician can be found to defend them. \hfill ---\citep[p. 193]{Senn:2001fn}
\end{quote}

Not only are P-values rarely defended, they are frequently derided (e.g. \cite{Berger:1987tf,Lecoutre:2001uy,Goodman:2001tx,Wagenmakers:2007ur}). Even so, support for the continued use of P-values for at least some purposes with some caveats can be found (e.g. \cite{Nickerson:2000th,Senn:2001fn,GarciaPerez:2016cr,Krueger:2017fz}). One crucial caveat is that a clear distinction has to be drawn between the dichotomisation of P-values into `significant' or `not significant' (typically on the basis of a threshold set at 0.05) and the evidential meaning of the actual numerically specified P-value. The former comes from a \textit{hypothesis test} and the latter from a \textit{significance test}. Contrary to what many readers will think and have been taught, they are not the same things. It might be argued that the battle to retain a clear distinction between significance tests and hypothesis tests has long been lost, but I have to continue that battle here because that distinction is critical for understanding the uses and misuses of P-values. Detailed accounts can also be found elsewhere \citep{Huberty:1993vx,Senn:2001fn,Hubbard:2003vq,Lenhard:2006aa,Hurlbert:2009ux,Lew:2012df}.

\subsection{Hypothesis test and Significance test}

When comparing significance tests and hypothesis tests it is conventional to note that the former are `Fisherian' (or, perhaps, ``neoFisherian'' \citep{Hurlbert:2009ux}) and the latter are `Neyman--Pearsonian'. R.A. Fisher did not invent significance tests per se---Gossett published what became Student's $t$-test before Fisher's career had begun \citep{Student:1908tf} and even that is not the first example---but Fisher did effectively popularise their use with his book \textit{Statistical Methods for Research Workers} (\citeyear{Fisher1925}), and he is credited with (or blamed for!) the convention of P$<0.05$ as a criterion for `significance'. It is important to note that Fisher's `significant' denoted something along the lines of worthy of further consideration or investigation, which is different to what is denoted by the same word applied tot he results of a hypothesis test. Hypothesis tests came later, with the 1933 paper by Neyman \& Pearson that set out the workings of dichotomising hypothesis tests and also introduced of the ideas ``errors of the first kind'' (false positive errors; type I errors) and ``errors of the second kind'' (false negative errors; type II errors) and a formalisation of the concept of statistical power.

A Neyman--Pearsonian hypothesis test is more than a simple statistical calculation. It is a method that properly encompasses experimental planning and experimenter behaviour as well. Before an experiment is conducted, the experimenter chooses  $\alpha$, the size of the critical region in the distribution of the test statistic, on the basis of the acceptable false positive (i.e. type I) error rate and sets the sample size on the basis of an acceptable false negative (i.e. type II) error rate. In effect the sample size, power\footnote{The `power' of the experiment is one minus the false positive error rate, but it is a function of the true effect size, as explained later.}, and $\alpha$ are traded off against each other to obtain an experimental design with the appropriate mix of cost and error rates. In order for the error rates of the procedure to be well calibrated, the sample size and $\alpha$ have to be set in advance of the experiment being performed, a detail that is often overlooked by pharmacologists. After the experiment has been run and the data are in hand, the mechanics of the test involves a determination of whether  the observed value of the test statistic lies within a pre-determined critical region under the sampling distribution provided by a statistical model and the null hypothesis. When the observed value of the test statistic falls within the critical range the result is `significant' and the analyst discards the null hypothesis. When the observed test statistic falls outside the critical range the result is `not significant' and the null hypothesis is not discarded.
 
 In current practice, dichotomisation of results into significant and not significant is most often made on the basis of the observed P-value being less than or greater than a conventional threshold of 0.05, so we have the familiar P$<0.05$ for $\alpha=0.05$. The one-to-one relationship between the test statistic being within the critical range and the P-value being less than $\alpha$ means that such practice is not intrinsically problematical, but using a P-value as an intermediate in a hypothesis test obscures the nature of the test and contributes to the conflation of significance tests and hypothesis tests.

 The classical Neyman--Pearsonian hypothesis test is an acceptance procedure, or a decision theory procedure \citep{Birnbaum:1977tv,Hurlbert:2009ux} that does not require, or provide, a P-value. Its output is a binary decision: either reject the null hypothesis or fail to reject the null hypothesis. In contrast, a Fisherian significance test yields a P-value that encodes the evidence in the data against the null hypothesis, but not, directly, a decision. The P-value is the probability of observing data as extreme as that observed, or more extreme, when the null hypothesis is true. That probability is generated or determined by a statistical model of some sort, and so we should really include the phrase `according to the statistical model' into the definition. In the Fisherian tradition\footnote{
It has been argued that because Fisher regularly described experimental results as `significant' or 'not significant' he was treating P-values dichotomously and that he used a fixed threshold for that dichotomisation (e.g. \citep[pp. 51--53]{LehmannFN}). However, Fisher meant the word `significant' to denote only that a result that is worthy of attention and follow up, and he quoted P-values as being less than 0.05, 0.02, and 0.01 because he was was working from tables of critical values of test statistics rather than laboriously calculating exact P-values manually. He wrote about the issue on several occasions, for example this:

\begin{quote}
Convenient as it is to note that a hypothesis is contradicted at some familiar level of significance such as 5\% or 2\% or 1\% we do not, in Inductive Inference, ever need to lose sight of the exact strength which the evidence has in fact reached, or to ignore the fact that with further trial it might come to be stronger, or weaker.
\hfill---\cite[p. 25]{FisherDOE}
\end{quote}
} a P-value is interpreted evidentially: the smaller the P-value the stronger the evidence against the null hypothesis and the more implausible the null hypothesis is, according to the statistical model. No behavioural or inferential consequences attach to the observed P-value and no threshold need be applied because the P-value is a continuous index.

In practice the probabilistic nature of P-values has proved difficult to use because people tend to mistakenly assume that the P-value measures the probability of the null hypothesis or the probability of an erroneous decision---it seems that they prefer any probability that is more noteworthy or less of a mouthful than the probability according to a statistical model of observing data as extreme or more extreme when the null hypothesis is true. Happily, there are no ordinary uses of P-values that require them to be interpreted as probabilities. My advice is to forget that P-values can be defined as probabilities and instead look at them as indices of surprisingness or unusualness of data: the smaller the P-value the more surprising are the data compared to what the statistical model predicts when the null hypothesis is true.

Conflation of significance tests and hypothesis tests may be encouraged by their apparently equivalent outputs (significance and P-values), but the conflation is too often encouraged by textbook authors, even to the extent of presenting a hybrid approach containing features of both. The problem has deep roots: when Neyman \& Pearson published their hypothesis test in 1933 it was immediately assumed that their test was an extension of Fisher's significance tests. Substantive differences in the philosophical and theoretical underpinnings soon became apparent to the protagonists and a long-lasting and bitter personal enmity developed between Fisher and Neyman \citep{Lenhard:2006aa, LehmannFN}. That feud seems likely to be one of the causes of the confusion that we have today as it has been suggested that authors of statistics textbooks avoided taking sides in the feud---an understandable response given vehemence and the forceful personalities of the protagonists---either by presenting only one of the approaches without mention of the other or by presenting a mixture of both \citep{CowlesBook1989,Huberty:1993vx,Halpin:2006wn}.

Whatever the origin of the confusion, the fact that significance tests and hypothesis tests are rarely explained as distinct alternatives in textbooks, encourages many to mistakenly assume that `significance test' and `hypothesis test' are synonyms. It also encourages the use a hybrid of the two which is commonly called NHST (for Null Hypothesis Significance Test). NHST has been derided, for example as an ``inconsistent mishmash'' \cite{Gigerenzer:1998wa} and as a ``jerry-built framework'' \citep[p. 1]{Krueger:2017fz} but versions of NHST are nonetheless more common than well-constructed hypothesis tests and significance tests together. Users of NHST almost universally assume that they are `doing it right' and the assumption that P-value equals NHST persists, largely unnoticed, particularly in the commentaries of those clamouring for the elimination of P-values. I therefore feel compelled to add to the list of derogatory epithets: NHST is like a reverso-platypus. The platypus was at one time derided as a fake\footnote{Well, that's the conventional wisdom, but it may be an exaggeration. The first scientific description of the ``duck-billed platypus'' was done in England by Shaw \& Nodder (\citeyear{Shaw1789}), who wrote ``Of all Mammalia yet known it seems the most extraordinary in its conformation; exhibiting the perfect resemblance of the beak of a Duck engrafted on the head of a quadruped. So accurate is the similitude that, at first view, it naturally excites the idea of some deceptive preparation by artificial means''. If Shaw \& Nodder really thought it a fake, they did not do so for long.}---a composite creature consisting of parts of several animals---but is a real animal, rare but beautiful, and perfectly adapted to its ecological niche. The common NHST is assumed by its many users to be a proper statistical procedure but is, in fact, an ugly composite, maladapted for almost all analytic purposes.

\subsection{Contradictory instructions}
\label{SecContradictory}

No-one should be using NHST, but should we use hypothesis testing or significance testing? The answer should depend on what your analytical objectives are, but in practice it more often depends on who you ask. Not all advice is good advice, and not even the experts agree. Responses to the American Statistical Association's official statement on P-values provides a case in point. In response to the widespread expressions of concern over the misuse and misunderstanding of P-values, the ASA convened a group of experts to consider the issues and to collaborate on drafting an official statement on P-values \citep{Wasserstein:2016jo}. Invited commentaries were published alongside the final statement, and even a brief reading of those commentaries on the statement will turn up misgivings and disagreements. Given that most of the commentaries were written by participants in the expert group, such disquiet and dissent confirms the difficulty of this topic. It should also should signal to readers that their practical familiarity with P-values does not ensure that they understand P-values.

The official ASA statement on P-values sets out six numbered principles concerning P-values and scientific inference:

\begin{enumerate}
	\item P-values can indicate how incompatible the data are with a specified statistical model.
	\item P-values do not measure the probability that the studied hypothesis is true, or the chance that the data were produced by random chance.
\item Scientific conclusions and business or policy decisions should not be based only on whether a P-value passes a specific threshold.
	\item Proper inference requires full reporting and transparency.
	\item A P-value, or statistical significance, does not measure the size of an effect or the importance of a result.
	\item By itself, a P-value does not provide a good measure of evidence regarding a model or hypothesis.
\end{enumerate}

Those principles are all sound---some derive directly from the definition of P-values and some are self-evidently good advice about the formation and reporting of scientific conclusions---but hypothesis tests and significance tests are not even mentioned in the statement and so it does not directly answer the question about whether we should use significance tests or hypothesis tests that I asked at the start of this section. Nevertheless, the statement offers a useful perspective and is not entirely neutral on the question. It urges against the use of a threshold in Principle 3 which says ``Scientific conclusions and business or policy decisions should not be based only on whether a p-value passes a specific threshold.'' Without a threshold we cannot use a hypothesis test. Lest any reader think that the intention is that P-values should not be used, I point out that the explanatory note for that principle in the ASA document begins thus:

\begin{quote}
Practices that reduce data analysis or scientific inference to mechanical ``bright-line'' rules (such as ``$p<0.05$'') for justifying scientific claims or conclusions can lead to erroneous beliefs and poor decision making. \hfill ---\citep[p. 131]{Wasserstein:2016jo}
\end{quote}

``Bright-line rule'' is an American legal phrase denoting an approach to simplifying ambiguous or complex legal issues by establishment of a clear, consistent ruling on the basis of objective factors. In other words, subtleties of circumstance and subjective factors are ignored in favour of consistency and simplicity. Such a rule might be useful in the legal setting, but it does not sound like an approach well-suited to the considerations that should underlie scientific inference. It is unfortunate, therefore, that a mechanical bright-line rule is so often used in basic pharmacological research, and even worse that it is demanded by the instructions to authors of the \textit{British Journal of Pharmacology}:

\begin{quote}
When comparing groups, a level of probability ($P$) deemed to constitute the threshold for statistical significance should be defined in Methods, and not varied later in Results (by presentation of multiple levels of significance). Thus, ordinarily $P < 0.05$ should be used throughout a paper to denote statistically significant differences between groups. \hfill---\citep{CurtisBJP}
\end{quote}

\begin{figure}
\begin{center}
\includegraphics[width=0.75\linewidth]{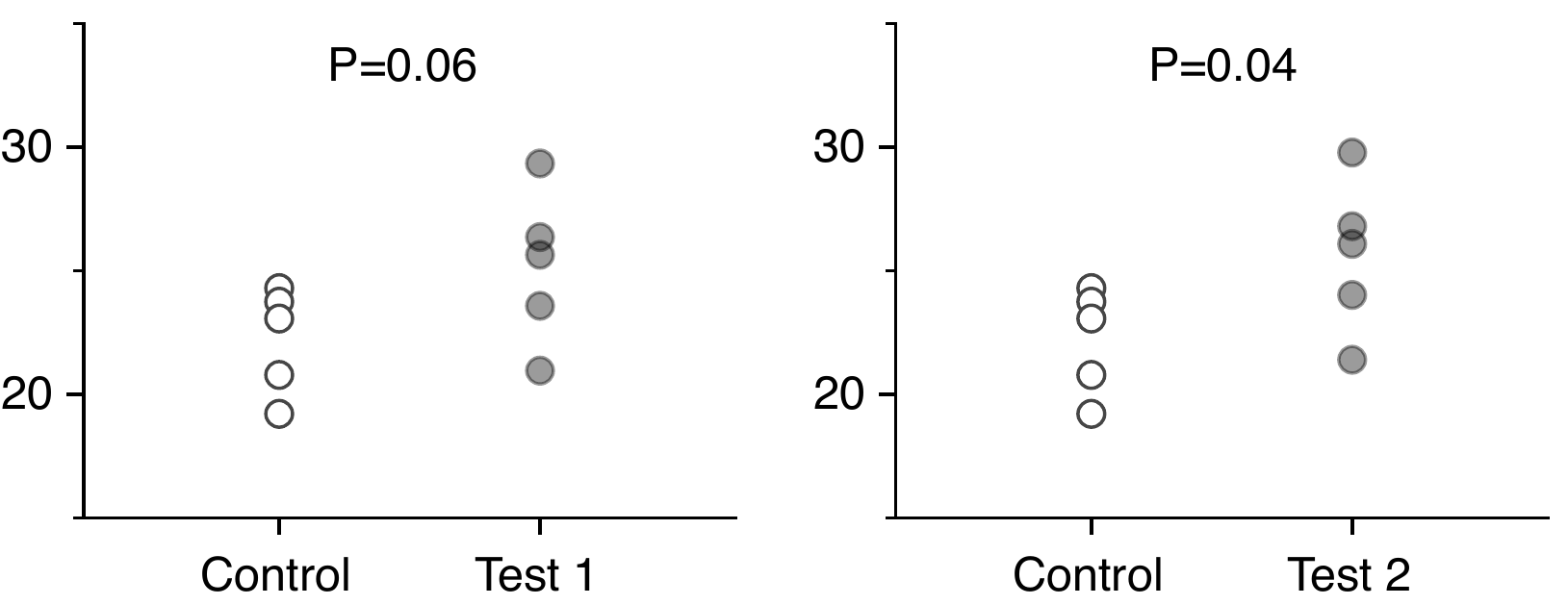}
\end{center}
\caption{\textit{P=0.04 is not very different from P=0.06.} Pseudo-data devised to yield one-tailed P=0.06 (left) and P=0.04 (right) from a Student's $t$-test for independent samples, $n=5$ per group. The y-axis is an arbitrarily scaled measure.}
\label{FigP0.04}
\end{figure}

An updated version of the guidelines retains those instructions \citep{Curtis:2018jc}, but because it is a bad instruction I present three objections. The first is that routine use of an arbitrary P-value threshold for declaring a result significant ignores almost all of the evidential content of the P-value by forcing an all-or-none distinction between a P-value small enough and one not small enough. The arbitrariness of a threshold for significance is well known and flows from the fact that there is no natural cutoff point or inflection point in the scale of P-values. Anyone who is unconvinced that it matters should note that the evidence in a result of P=0.06 is not so different from that in a result of P=0.04 as to support an opposite conclusion (Figure \ref{FigP0.04}).

The second objection to the instruction to use a threshold of P$<0.05$ is that exclusive focus on whether the result is above or below the threshold blinds analysts to information beyond the sample in question. If the statistical procedure says discard the null hypothesis (or don't discard it) then that statistical decision seems to override and make redundant any further considerations of evidence, theory, or scientific merit. That is quite dangerous, because all relevant material should be considered and integrated into scientific inferences.

The third objection refers to the strength of evidence needed to reach the threshold: the \textit{British Journal of Pharmacology} instruction licenses claims on the basis of relatively weak evidence.\footnote{Accepting P=0.05 as a sufficient reason to suppose that a treatment is effective is akin to accepting 50\% as a passing grade: it's traditional in many settings, but it's far from reassuring.} The evidential disfavouring of the null hypothesis in a P-value close to 0.05 is surprisingly weak when viewed as a likelihood ratio or Bayes factor \citep{Goodman:1988ui, Johnson:2013eh, Benjamin:2018gh}, a weakness that can be confirmed by simply `eyeballing' Figure \ref{FigP0.04}. 

A fixed threshold corresponding to weak evidence might sometimes be reasonable, but often it is not. As Carl Sagan said: ``Extraordinary claims require extraordinary evidence.''\footnote{That phrase comes from the television series \textit{Cosmos}, 1980, but may derive from Laplace (\citeyear{Laplace1812}), who wrote ``The weight of evidence for an extraordinary claim must be proportioned to its strangeness.'' [translated, the original is in French].} It would be possible to overcome this last objection by setting a lower threshold whenever an extraordinary claim is to be made, but the \textit{British Journal of Pharmacology} instructions preclude such a choice by insisting that the same threshold be applied to all tests within the whole study.   

There has been a serious proposal that a lower default threshold of P$<0.005$ be adopted as the default \citep{Johnson:2013eh,Benjamin:2018gh}, but even if that would ameliorate the weakness of evidence objection, it doesn't address all of the problems posed by dichotomising results into significant and not significant, as is acknowledged by the many authors of that proposal. 

 Should the \textit{British Journal of Pharmacology} enforce its guideline on the use of Neyman--Pearsonian hypothesis testing with a fixed threshold for statistical significance? Definitely not, and laboratory pharmacologists should usually avoid them because the nature those tests is ill-suited to the reality of basic pharmacological studies.
 
 The shortcoming of hypothesis testing is that it offers an all-or-none outcome and it engenders a one-and-done response to an experiment. All-or-none in that the significant or not significant outcome is dichotomous. One-and-done because once a decision has been made to reject the null hypothesis there is little apparent reason to re-test that null hypothesis the same way, or differently. There is no mechanism within the classical Neyman--Pearsonian hypothesis testing framework for a result to be treated as provisional. That is not particularly problematical in the context of a classical randomised clinical trial (RCT) because an RCT is usually conducted only after preclinical studies have addressed the relevant biological questions. That allows the scientific aims of the study to be simple---they are designed to provide a definitive answer to the primary question. An all-or-none one-and-done hypothesis test is therefore appropriate for an RCT.\footnote{Clinical trials are sometimes aggregated in meta-analyses, but the substrate for meta-analytical combination is the observed effect sizes and sample sizes of the individual trials, not the dichotomised significant or not significant outcomes.} But the majority of basic pharmacological laboratory studies do not have much in common with an RCT because they consist of a series of interlinked and inter-related experiments contributing variously to the primary inference. For example, a basic pharmacological study will often include experiments that validate experimental methods and reagents, concentration-response curves for one or more of drugs, positive and negative controls, and other experiments subsidiary to the main purpose of the study. The design of the `headline' experiment (assuming there is one) and interpretation of its results is dependent on the results of those subsidiary experiments, and even when there is a singular scientific hypothesis, it might be tested in several ways using observations within the study. It is the aggregate of all of the experiments that inform the scientific inferences. The all-or-none one-and-done outcome of a hypothesis test is less appropriate to basic research than it is to a clinical trial.
 
 Pharmacological laboratory experiments also differ from RCTs in other ways that are relevant to the choice of statistical methodologies. Compared to an RCT, basic pharmacological research is very cheap, the experiments can be completed very quickly, with the results available for analysis almost immediately. Those advantages mean that a pharmacologist might design some of the experiments within a study in response to results obtained in that same study,\footnote{Yes, that is also done in `adaptive' clinical trials, but they are not the archetypical RCT that is the comparator here.} and so a basic pharmacological study will often contain preliminary or exploratory research. Basic research and clinical trials also differ in the consequences of erroneous inference. A false positive in an RCT might prove very damaging by encouraging the adoption of an ineffective therapy, but in the much more preliminary world of basic pharmacological research a false positive result might have relatively little influence on the wider world. It could be argued that statistical protections against false positive outcomes that are appropriate in the realm of clinical trials can be inappropriate in the realm of basic research. This idea is illustrated in a later section of this chapter.
 
The multi-faceted nature of the basic pharmacological study means that statistical approaches yielding dichotomous yes or no outcomes are less relevant than they are to the archetypical RCT. The scientific conclusions drawn from basic pharmacological experiments should be based on thoughtful consideration of the entire suite of results in conjunction with any other relevant information, including both pre-existing evidence and theory.  The dichotomous all-or-none, one-and-done hypothesis test is poorly adapted to the needs of basic pharmacological experiments, and is probably poorly adapted to the needs of most basic scientific studies. Scientific studies depend on a detailed evaluation of evidence but a hypothesis test does not fully support such an evaluation.
 
 \subsection{Evidence is local; error rates are global}
 
A way to understand difference between the Fisherian significance test and the Neyman--Pearsonian hypothesis test is to recognise that the former supports `local'  inference, whereas the latter is designed to protect against `global' long-run error. The P-value of a significance test is local because it is an index of the evidence in \textit{this} data against \textit{this} null hypothesis. In contrast, the hypothesis test decision regarding rejection of the null hypothesis is global because it is based on a parameter, $\alpha$, which is set without reference to the observed data. The long run performance of the hypothesis test is a property of the procedure itself and is independent of any particular data, and so it is global. Local evidence; global errors. This is not an ahistoric imputation, because Neyman \& Pearson were clear about their preference for global error protection rather than local evidence and their objectives in devising hypothesis tests:

\begin{quote}
We are inclined to think that as far as a particular hypothesis is concerned, no test based upon the theory of probability can by itself provide any valuable evidence of the truth or falsehood of that hypothesis.

But we may look at the purpose of tests from another view-point. Without hoping to know whether each separate hypothesis is true or false, we may search for rules to govern our behaviour with regard to them, in following which we insure that, in the long run of experience, we shall not be too often wrong. 

\hfill---\cite{Neyman:1933ud}
\end{quote}

The distinction between local and global properties or information is relatively little known, but Liu \& Meng \citeyear{LiuMeng2016} offer a much more technical and complete discussion of the local/global distinction, using the descriptors `individualised' and `relevant' for the local and the `robust' for the global. They demonstrate a trade-off between relevance and robustness that requires judgement on the part of the analyst. In short, the desirability of methods that have good long-run error properties is undeniable, but paying attention exclusively to the global blinds us to the local information that is relevant to inferences. The instructions of the \textit{British Journal of Pharmacology} are inappropriate because they attend entirely to the global and because the dichotomising of each experimental result into significant and not significant hinders thoughtful inference. 

Many of the battles and controversies regarding statistical tests swirl around issues that might be clarified using the local versus global distinction, and so it will be referred to repeatedly in what follows.

\subsection{On the scaling of P-values}

\begin{figure}
\begin{center}
\includegraphics[width=0.5\linewidth]{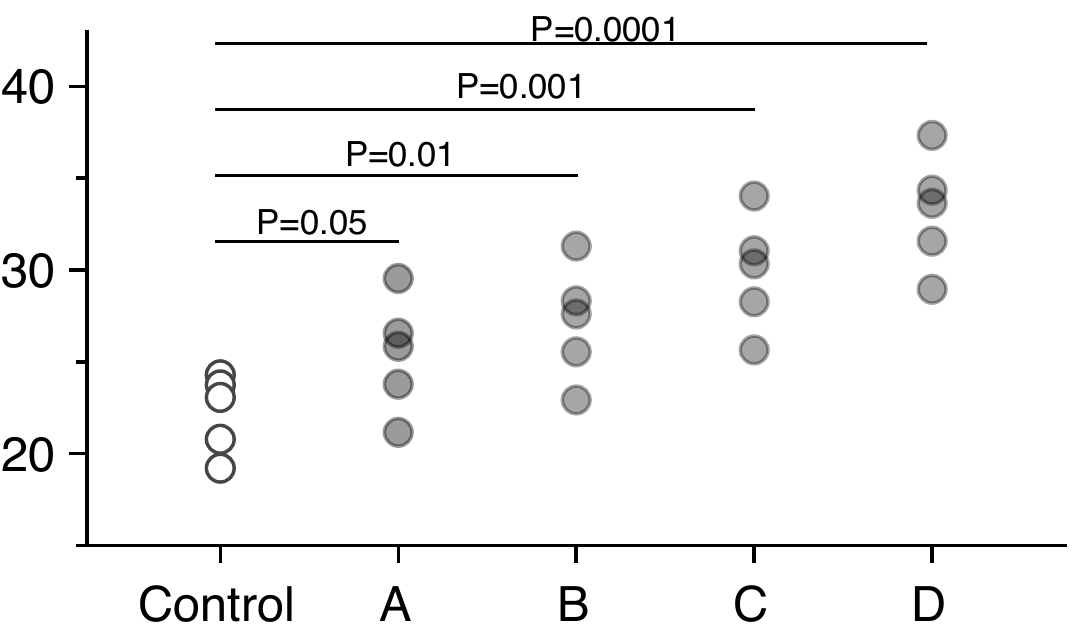}
\end{center}
\caption{
\textit{What simple evidence looks like.} Pseudo-data devised to yield one-tailed P-values from 0.05 to 0.0001 from a Student's $t$-test for independent samples, $n=5$ per group. The left-most group of values is the control against which each of the other sets is compared, and the pseudo-datasets A, B, C, and D were generated by arithmetic adjustment of a single dataset to obtain the indicated P-values. The y-axis is an arbitrarily scaled measure.
}
\label{FigVariousPs}
\end{figure}

\begin{figure}
\begin{center}
\includegraphics[width=0.5\linewidth]{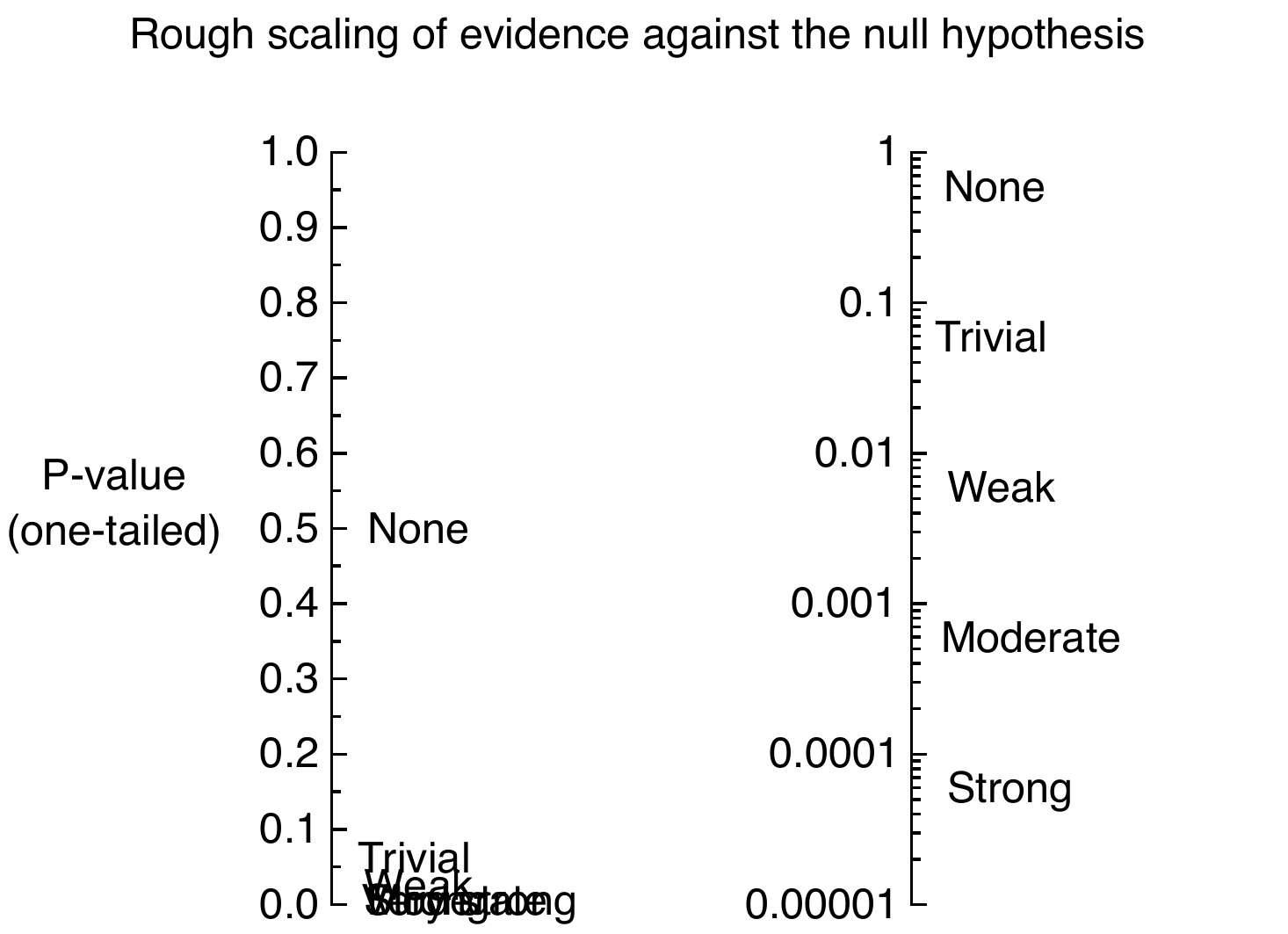}
\end{center}
\caption{
\textit{Evidential descriptors for P-values.} Strength of evidence against the null hypothesis scales semi-geometrically with the smallness of the P-value. Note that the descriptors for strength of evidence are illustrative only, and it would be a mistake to assume, for example, that a P-value of 0.001 indicates moderately strong evidence against the null hypothesis in every circumstance.
}
\label{FigPvsEv}
\end{figure}

In order to be able to safely interpret the local, evidential, meaning of a P-value, a pharmacologist should understand its scaling. Just like the $\textup{EC}_{50}$s with which pharmacologists are so familiar, P-values have a bounded scale, and just as is the case with $\textup{EC}_{50}$s it makes sense to scale P-values geometrically (or logarithmically). The non-linear relationship between P-values and an intuitive scaling of evidence against the null hypothesis can be gleaned from Figure \ref{FigVariousPs}. Of course, a geometric scaling of the evidential meaning of P-values implies that the descriptors of evidence should be similarly scaled and so such a scale is proposed in Figure \ref{FigPvsEv}, with P-values around 0.05 being called `trivial' in recognition of the relatively unimpressive evidence for a real difference between condition A and control in Figure \ref{FigVariousPs}.

Attentive readers will have noticed that the P-values in Figures \ref{FigP0.04}, \ref{FigVariousPs}, and \ref{FigPvsEv} are all one-tailed. The number of tails that published P-values have is inconsistent, is often unspecified, and the number of tails that a P-value \textit{should} have is controversial (e.g. see \cite{Dubey:1991uh,Bland:1994jg,Kobayashi:1997uj,Freedman:2008ck,Lombardi:2009fc,Ruxton:2010jx}). Arguments about P-value tails are regularly confounded by differences between local and global considerations. The most compelling reasons to favour two tails relate to global error rates, which means that they apply only to P-values that are dichotomised into significant and not significant in a hypothesis test. Those arguments can safely be ignored when P-values are used as indices of evidence and I therefore recommend one-tailed P-values for general use in pharmacological experiments---as long as the P-values are interpreted as evidence and not as a surrogate for decision. (Either way, the number of tails should always be specified.)

 \subsection{Power and expected P-values}
 
 The Neyman--Pearsonian hypothesis test is a decision procedure that, with a few assumptions, can be an optimal procedure. Optimal only in the restricted sense that the smallest sample gives the highest power to reject the null hypothesis when it is false, for any specified rate of false positive errors. To achieve that optimality the experimental sample size and $\alpha$ are selected prior to the experiment using a power analysis and with consideration of the costs of the two specified types of error and the benefits of potentially correct decisions. In other words, there is a loss function built into the design of experiments. However, outside of the clinical trials arena, few pharmacologists seem to design experiments in that way. For example, a study of 22 basic biomedical research papers published in \textit{Nature Medicine} found that none of them included any mention of a power analysis for setting the sample size \citep{Strasak:2007vj}, and a simple survey of the research papers in the most recent issue of \textit{British Journal of Pharmacology} (2018, issue 17 of volume 175) gives a similar picture with power analyses specified in only one out the 11 research papers that used P $<$ 0.05 as a criterion for statistical significance. It is notable that all of those \textit{BJP} papers included statements in their methods sections claiming compliance with the guidelines for experimental design and analysis, guidelines that include this as the first key point:

\begin{quote}
Experimental design should be subjected to ‘a priori power analysis’ so as to ensure that the size of treatment and control groups is adequate[\dots] \hfill---\citep{CurtisBJP}
\end{quote}

\noindent The most recent issue of \textit{Journal of Pharmacology and Experimental Therapeutics} (2018, issue 3 of volume 366) similarly contains no mention of power of sample size determination in any of its 9 research papers, although none of its authors had to pay lip service to guidelines requiring it. 

In reality, power analyses are not always necessary or helpful. They have no clear role in the design of a preliminary or exploratory experiment that is concerned more with hypothesis generation than hypothesis testing, and a large fraction of the experiments published in basic pharmacological journals are exploratory or preliminary in nature. Nonetheless, they are described here in detail because experience suggests they are mysterious to many pharmacologists and they are very useful for planning confirmatory experiments.

For a simple test like Student's $t$-test a pre-experiment power analysis for determination of sample size is easily performed. The power of a Student's $t$-test is dependent on: (i) the predetermined acceptable false positive error rate, $\alpha$ (bigger $\alpha$ gives more power); (ii) the true effect size, which we will denote as $\delta$ (more power when $\delta$ is larger); (iii) the population standard deviation, $\sigma$ (smaller $\sigma$ gives more power); and (iv) the sample size (larger $n$ for more power). The common approach to a power test is to specify an effect size of interest and the minimum desired power, so say we wish to detect a true effect of $\delta=3$ in a system where we expect the standard deviation to be $\sigma=2$. The free software\footnote{www.r-project.org} called R has the function power.t.test() that gives this result:

\begin{verbatim}
	 > power.t.test(delta=3, sd=2, power=0.8, sig.level = 0.05, 
	 alternative ='one.sided', n=NULL)

     Two-sample t test power calculation 

              n = 6.298691
          delta = 3
             sd = 2
      sig.level = 0.05
          power = 0.8
    alternative = one.sided

NOTE: n is number in *each* group
\end{verbatim}

\noindent It is conventional to round the sample size up to the next integer so the sample size would be 7 per group.

\begin{figure}
\begin{center}
\includegraphics[width=0.75\linewidth]{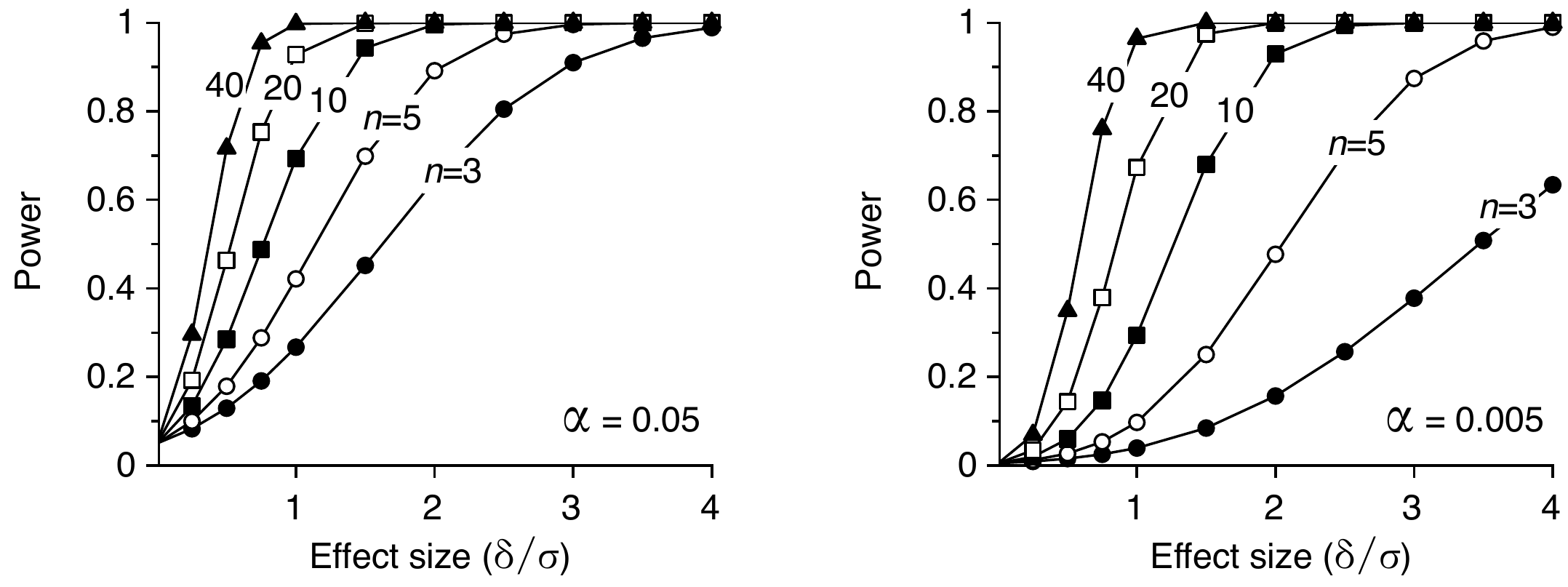}
\end{center}
\caption{
\textit{Power functions for $\alpha=0.05$ and $0.005$.} Power of one-sided Student's $t$-test for independent samples expressed as a function of standardised true effect size $\delta/\sigma$ for sample sizes (per group) from $n=3$ to $n=40$. Note that $\delta = \mu_1-\mu_2$ and $\sigma$ are population parameters rather than sample estimates.}
\label{FigPowerFns}
\end{figure}

While a single point power analysis like that is straightforward, it provides relatively little information compared to the information supplied by the analyst, and its output is specific to the particular effect size specified, an effect size that more often than not has to be `guesstimated' instead of estimated because it is the unknown that is the object of study. A plot of power versus effect size is far more informative than the point value supplied by the conventional power test (Figure \ref{FigPowerFns}). Those graphical power functions show clearly the three-way relationship between sample size, effect size and the risk of a false negative outcome (i.e. one minus the power).
 
Some experimenters are tempted to perform a post-experiment power analysis when their observed P-value is unsatisfyingly large. They aim to answer the question of how large the sample \textit{should} have been, and proceed to plug in the observed effect size and standard deviation and pulling out a larger sample size---always larger---that might have given them the desired small P-value. Their interpretation is then that the result \textit{would have been significant} but for the fact that the experiment was underpowered. That interpretation ignores that fact that the observed effect size might be an exaggeration, or the observed standard deviation might be an underestimation and the null hypothesis might be true! Such a procedure is generally inappropriate and dangerous \citep{Hoenig:2001vq}. There is a one to one correspondence of observed P-value and post-experiment power and no matter what the sample size, a larger than desired P-value \textit{always} corresponds to a low power at the observed effect size, whether the null hypothesis is true or false. Power analyses are useful in the design of experiments, not for the interpretation of experimental results.

\begin{figure}
\begin{center}
\includegraphics[width=0.75\linewidth]{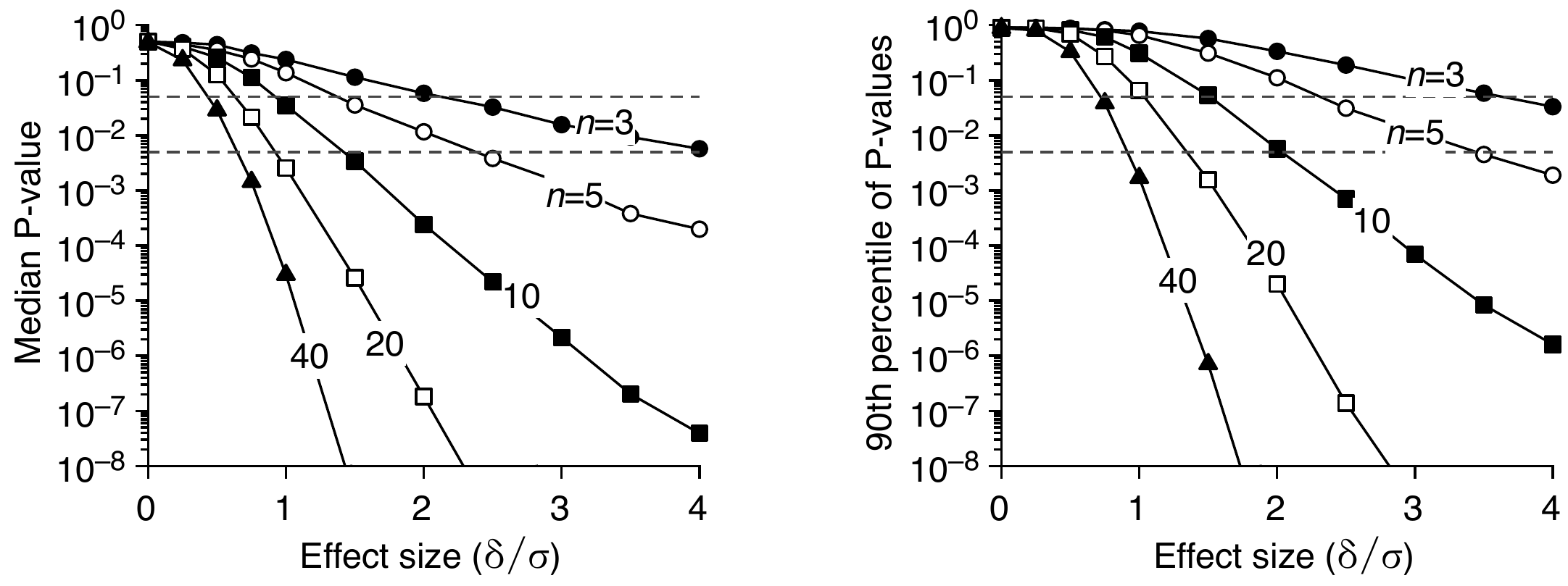}
\end{center}
\caption{
\textit{Expected P-value functions} P-values expected from Student's $t$-test for independent samples expressed as a function of standardised true effect size $\delta/\sigma$ for sample sizes (per group) from $n=3$ to $n=40$. The graph on the left shows the median of expected P-values (i.e. the 50th percentile) and the graph on the right shows the 90th percentile. It can be expected that 50\% of observed P-values will lie below the median lines and 90\% will lie below the 90th percentile lines for corresponding sample sizes and effect sizes. The dashed lines indicate P=0.05 and 0.005.}

\label{FigPvalFns}
\end{figure}

Power analyses are tied closely to dichotomising Neyman--Pearsonian hypothesis tests, even when expanded to provide full power functions as in Figure \ref{FigPowerFns}. However, there is an alternative more closely tied to Fisherian significance testing---an approach better aligned to the objectives of evidence gathering. That alternative is a plot of average expected P-values as functions of effect size and sample size \citep{Sackrowitz:1999uu,Bhattacharya:2002kra}. The median is more relevant than the mean, both because the distribution of expected P-values is very skewed and because the median value offers a convenient interpretation of there being a 50:50 bet that and observed P-value will be either side of it. An equivalent plot showing the 90th percentile of expected P-values gives another option for experiment sample size planning purposes (Figure \ref{FigPvalFns}).

Should the \textit{British Journal of Pharmacology} enforce its power guideline? In general no, but pharmacologists should use power curves or expected P-value curves for designing some of their experiments, and ought to say so when they do. Power analyses for sample size are very important for experiments that are intended to be definitive and decisive, and that's why sample size considerations are dealt with in detail when planning clinical trials. Even though the majority of experiments in basic pharmacological research papers are not like that, as discussed above, even preliminary experiments should be planned to a degree, and power curves and expected P-value curves are both useful in that role.

\section{Practical problems with P-values}

The sections above deal with the most basic misconceptions regarding the nature of P-values, but critics of P-values usually focus on other important issues. In this section I will deal with the significance filter, multiple comparisons, and some forms of P-hacking, and I need to point out immediately that most of the issues are not specific to P-values even if some of them are enabled by the unfortunate dichotomisation of P-values into significant and not significant. In other words, the practical problems with P-values are largely the practical problems associated with the \textit{mis}use of P-values and with sloppy statistical inference generally.

\subsection{The significance filter exaggeration machine}

It is natural to assume that the effect size observed in an experiment is a good estimate of the true effect size, and in general that can be true. However, there are common circumstances where the observed effect size consistently overestimates the true, sometimes wildly so. The overestimation depends on the facts that experimental results exaggerating the true effect are more likely to be found statistically significant, and that we pay more attention to the significant results and are more likely to report them. The key to the effect is selective attention to a subset of results---the significant results---and so the process is appropriately called \textit{the significance filter}.

If there is nothing assume nothing untoward in the sampling mechanism,\footnote{That is not a safe assumption, in particular because a haphazard sample is not a random sample. When was the last time that you used something like a random number generator for allocation of treatments?} sample means are unbiassed estimators of population means and sample-based standard deviations are nearly unbiassed estimators of population standard deviations.\footnote{The variance is unbiassed but the non-linear square root transformation into the standard deviation damages that unbiassed-ness. Standard deviations calculated from small samples are biassed toward underestimation of the true standard deviation. For example, if the true standard deviation is 1 the expected average observed standard deviation for samples of $n=5$ is 0.92.} Because of that we can assume that, on average, a sample mean provides a sensible `guesstimate' for the population parameter and, to a lesser degree, so does the observed standard deviation. That is indeed the case for averages over all samples, but it cannot be relied upon for any particular sample. If attention has been drawn to a sample on the basis that it is `statistically significant' then that sample is likely to offer an exaggerated picture of the true effect. The phenomenon is usually called \textit{the significance filter}. The way it works is fairly easily described but, as usual, there are some complexities in its interpretation.

\begin{figure}
\begin{center}
\includegraphics[width=0.75\linewidth]{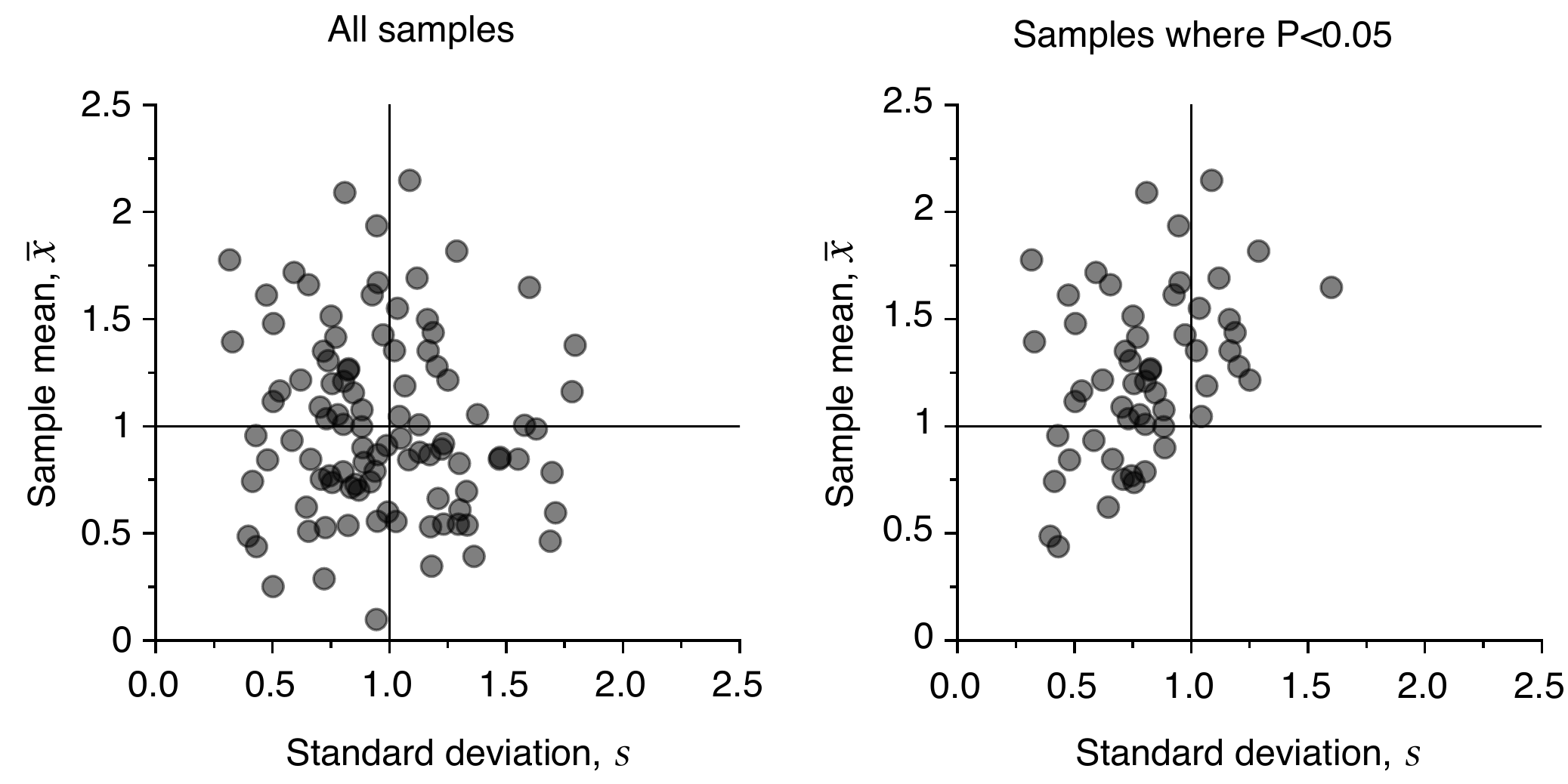}
\end{center}
\caption{
\textit{The significance filter.} The dots in the graphs are means and standard deviations of samples of $n=5$ drawn from a normally distributed population with mean $\mu=1$ and standard deviation $\sigma=1$. The left panel shows all 100 samples and the right panel shows only the results where P$<$0.05. The vertical and horizontal lines indicate the true parameter values. `Significant' results tend to over-estimate the population mean and under-estimate the population standard deviation.
}
\label{FigShotgun}
\end{figure}

Say we are in the position to run an experiment 100 times with random samples of $n=5$ from a single normally distributed population with mean $\mu=1$ and standard deviation $\sigma=1$. We would expect that, on average, the sample means, $\bar{x}$ would be scattered symmetrically around the true value of 1, and the sample-based standard deviations, $s$, would be scattered around the true value of 1, albeit slightly asymmetrically. A set of 100 simulations matching that scenario show exactly that result (see the left panel of Figure \ref{FigShotgun}), with the median of $\bar{x}$ being 0.97 and the median of $s$ being 0.94, both of which are close to the expected values of exactly 1 and about 0.92, respectively. If we were to pay attention only to the results where the observed P-value was less than 0.05 (with the null hypothesis being that the population mean is 0), then we get a different picture because the values are very biassed (see the right panel of Figure \ref{FigShotgun}). Among the `significant' results the median sample mean is 1.2 and the median standard deviation is 0.78.

The systematic bias of mean and standard deviation among `significant' results in those simulations might not seem too bad, but it is conventional to scale the effect size as the standardised ratio $\bar{x}/s$,\footnote{That ratio is often called Cohen's $d$. Pharmacologists should pay no attention to Cohen's specifications of small, medium and large effect sizes \citep{COHEN:1992wu} because they are much smaller than the effects commonly seen in basic pharmacological experiments.} and the median of that ratio among the `significant' results is fully 50\% larger than the correct value. What's more, the biasses get worse with smaller samples, with smaller true effect sizes, and with lower P-value thresholds for `significance'.

It is notable that even the results with the most extreme exaggeration of effect size in Figure \ref{FigShotgun}---550\%---would not be counted as an error within the Neyman--Pearsonian hypothesis testing framework! It would not lead to the false rejection of a true null or to an inappropriate failure to reject a false null and so it is neither a type I nor a type II error. But it is some type of error, a substantial error in estimation of the magnitude of the effect. The term \textit{type M error} has been devised for exactly that kind of error \citep{Gelman:2014cd}. A type M error might be underestimation as well as the overestimation, but overestimation is the more common in theory \citep{LuQiuDeng2018} and in practice \citep{Camerer:2018de}.

The effect size exaggeration coming from the significance filter is not a result of sampling, or of significance testing, or of P-values. It is a result of paying extra attention to a subset of all results---the `significant' subset.

The significance filter presents a peculiar difficulty. It leads to exaggeration \textit{on average}, but any particular result may well be close to the correct size whether it is `significant' or not. A real-world sample mean of, say, $\bar{x}=1.5$ might be an exaggeration of $\mu=1$, it might be an underestimation of $\mu=2$, or it might be pretty close to $\mu=1.4$ and there would be no way to be certain without knowing $\mu$, and if $\mu$ were known then the experiment would probably not have been necessary in the first place. That means that the possibility of a type M error looms over any experimental result that is interesting because of a small P-value, and that is particularly true when the sample size is small. The only way to gain more confidence that a particular significant result closely approximates the true state of the world is to repeat the experiment--the second result would not have been run through the significance filter and so its results would not have a greater than average risk of exaggeration and the overall inference can be informed by both results. Of course, experiments intended to repeat or replicate an interesting finding should take the possible exaggeration into account by being designed to have higher power than the original.

\subsection{Multiple comparisons}

\begin{figure}
\begin{center}
\includegraphics[width=0.6\linewidth]{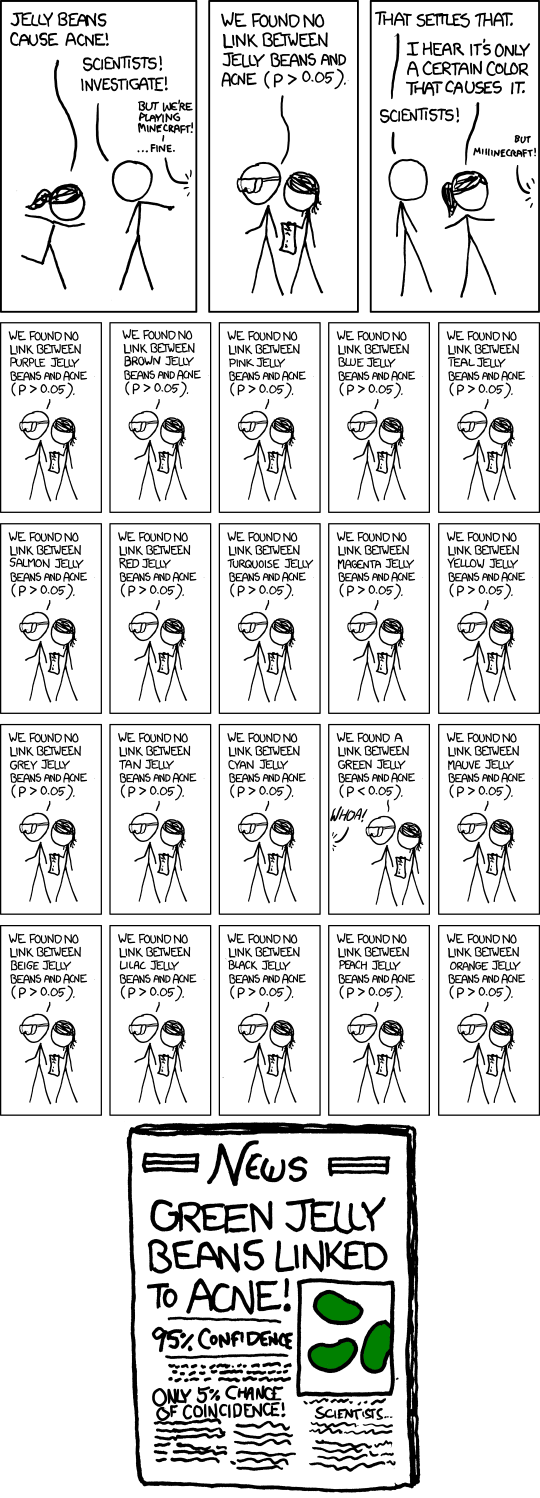}
\end{center}
\caption{
Multiple testing cartoon from XKCD, https://xkcd.com/882/ 
}
\label{FigXKCD}
\end{figure}

Multiple testing is the situating where the tension between global and local considerations are most stark. It is also the situation where the well-known jelly beans cartoon from XKCD.com is irresistable (Figure \ref{FigXKCD}). The cartoon scenario is that jelly beans were suspected of causing acne, but a test found ``no link between jelly beans and acne (P$>0.05$)", and so the possibility that only a certain colour of jelly bean causes acne is then entertained. All 20 colours of jelly bean are independently tested, with only the result from green jelly beans being significant, ``(P$<0.05$)''. The newspaper headline at the end of the cartoon mentions only the green jelly beans result, and it does that with exaggerated certainty. The usual interpretation of that cartoon is that the significant result with green jelly beans is likely to be a false positive because, after all, hypothesis testing with the threshold of P$<0.05$ is expected to yield a false positive one time in 20, on average, when the null is true.

 The more hypothesis tests there are, the higher the risk that one of them will yield a false positive result. The textbook response to multiple comparisons is to introduce `corrections' that protect an overall maximum false positive error rate by adjusting the threshold according to the number of tests in the family to give protection from inflation of the family-wise false positive error rate. The Bonferroni adjustment is the best-known method, and while there are several alternative `corrections' that perform a little better, none of those is nearly as simple. A Bonferroni adjustment for the family of experiments in the cartoon would preserve an overall false positive error rate of 5\% by setting a threshold for significance of $0.05/20= 0.0025$ in each of the 20 hypothesis tests.\footnote{You may notice that the first test of jelly beans without reference to colour has been ignored here. There is no set rule for saying exactly which experiments constitute a family for the purposes of correction of multiplicity.} It must be noted that such protection does not come for free, because adjustments for multiplicity invariably strip statistical power from the analysis.

We do not know whether the `significant' link between green jelly beans and acne would survive a Bonferroni adjustment because the actual P-values were not supplied,\footnote{That serves to illustrate one facet of the inadequacy of reporting `P less thans' in place of actual P-values.} but as an example, a P-value of 0.003, low enough to be quite encouraging as the result of a significance test, would be `not significant' according to the Bonferroni adjustment. Such a result that would present us with a serious dilemma because the inference supported by the local evidence would be apparently contradicted by global error rate considerations. However, that contradiction is not what it seems because the null hypothesis of the significance test P-value is a different null hypothesis from that tested by the Bonferroni-adjusted hypothesis test. The significance test null concerns only the green jelly beans whereas the null hypothesis of the Bonferroni is an omnibus null hypothesis that says that the link between green jelly beans on acne is zero \textit{and} the link between purple jelly beans on acne is zero \textit{and} the link between brown jelly beans is zero, and so on. The P-value null hypothesis is local and the omnibus null is global. The global null hypothesis might be appropriate before the evidence is available (i.e. for power calculations and experimental planning), but after the data are in hand the local null hypothesis concerning just the green jelly beans gains importance.

It is important to avoid being blinded to the local evidence by a non-significant global. After all, the pattern of evidence in the cartoon is \textit{exactly} what would be expected if the green colouring agent caused acne: green jelly beans are associated with acne but the other colours are not. (The failure to see an effect of the mixed jelly beans in the first test is easily explicable on the basis of the lower dose of green.) If the data from the trial of green jelly beans is independent of the data from the trials of other colours, then there is no way that the existence of those other data---or their analysis---can influence the nature of the green data. The green jelly bean data cannot logically have been affected by the fact that mauve and beige jelly beans were tested at a later point in time---the subsequent cannot affect the previous---and the experimental system would have to be bizarrely flawed for the testing of the purple or brown jelly beans to affect the subsequent experiment with green jelly beans. If the multiplicity of tests did not affect the data then it is only reasonable to say that it did not affect the evidence.

The omnibus global result does not cancel the local evidence, or even alter it, and yet the elevated risk of a false positive error is real. That presents us with a dilemma and, unfortunately, statistics does not provide a way around it. Global error rates and local evidence operate in different logical spaces \citep{ThompsonBook} and so there can be no strictly statistical way to weigh them together. All is not lost, though, because statistical limitations do not preclude thoughtful integration of local and the global issues when making inferences. We just have to be more than normally cautious when the local and global pull in different directions. For example, in the case of the cartoon, the evidence in the data favour the idea that green jelly beans are linked with acne (and if we had an exact P-value then we could specify the strength of favouring) but because the data were obtained by a method with a substantial false positive error rate we should be somewhat reluctant to take that evidence at face value. It would be up to the scientist in the cartoon (the one with safety glasses) to form a provisional scientific conclusion regarding the effect of green jelly beans, even if that inference is that any decision  should be deferred until more evidence is available. Whatever the inference, the evidence, theory, the method, any other corroborating or rebutting information should all be considered and reported.

\begin{quote}
A man or woman who sits and deals out a deck of cards repeatedly will eventually get a very unusual set of hands. A report of unusualness would be taken differently if we knew it was the only deal made, or one of a thousand deals, or one of a million deals, etc. \hfill---\cite[p. 133]{Tukey:1991vp}
\end{quote}

In isolation the cartoon experiments are probably only sufficient to suggest that the association between green jelly  acne is worthy of further investigation (with the earnestness of that suggestion being inversely related to the size of the relevant P-value). The only way to be in a position to report an inference concerning those jelly beans without having to hedge around the family-wise false positive error rate and the significance filter is to re-test the green jelly beans. New data from a separate experiment will be free from the taint of elevated family-wise error rates and untouched by the significance filter exaggeration machine. And, of course, \textit{all} of the original experiments should be reported alongside the new, as well as reasoned argument incorporating corroborating or rebutting information and theory.

The fact that a fresh experiment is necessary to allow a straightforward conclusion about the effect of the green jelly beans means that the experimental series shown in the cartoon is a preliminary, exploratory study. Preliminary or exploratory research is essential to scientific progress and can merit publication as long as it is reported completely and openly as preliminary. Too often scientists fall into the pattern of misrepresenting the processes that lead to their experimental results, perhaps under the mistaken assumption that science has to be hypothesis driven \citep{Medawar:1963,duPrel:2009kt,Howitt:2014kg}. That misrepresentation may take the form of a suggestion, implied or stated, that the green jelly beans were the intended subject of the study, a behaviour described as \textit{HARKing} for \textit{h}ypothesising \textit{a}fter the \textit{r}esults are \textit{k}nown, or \textit{cherry picking} where only the significant results are presented.  The reason that HARKing is problematical is that hypotheses cannot be tested using the data that suggested the hypothesis in the first place because those data \textit{always} support that hypothesis (otherwise they would not be suggesting it!), and cherry picking introduces a false impression of the nature of the total evidence and allows the direct introduction of experimenter bias. Either way, focussing on just the unusual observations from a multitude is bad science. It takes little effort and few words to say that 20 colours were tested and only the green yielded a statistically significant effect, and a scientist can (should) then hypothesise that green jelly beans cause acne and test that hypothesis with new data.

\subsection{P-hacking}

P-hacking is where an experiment or its analysis are directed at obtaining a small enough P-value to claim significance instead of being directed at the clarification of a scientific issue or testing of a hypothesis. Deliberate P-hacking does happen, perhaps driven by the incentives built into the systems of academic reward and publication imperatives, but most P-hacking is accidental---honest researchers doing `the wrong thing' through ignorance. P-hacking is not always as wrong as might be assumed, as the idea of P-hacking comes from paying attention exclusively to global consideration of error rates, and most particularly to false positive error rates. Those most stridently opposed to P-hacking will point to the increased risk of false positive errors, but rarely to the lowered risk of false negative errors. I will recklessly note that some categories of P-hacking look entirely unproblematical when viewed through the prism of local evidence. The local versus global distinction allows a more nuanced response to P-hacking.

Some P-hacking is outright fraud. Consider this example that has recently come to light:

\begin{quote}
One sticking point is that although the stickers increase apple selection by 71\%, for some reason this is a p value of $.06$. It seems to me it should be lower. Do you want to take a look at it and see what you think. If you can get the data, and it needs some tweeking, it would be good to get that one value below .05.

---Email from Brian Wansink to David Just on Jan. 7, 2012.	\citep{BuzzFeedWansink}
\end{quote}

\noindent  I do not expect that any readers would find P-hacking of that kind to be acceptable. However, the line between fraudulent P-hacking and the more innocent P-hacking through ignorance is hard to define, particularly so given the fact that some behaviours derided as P-hacking can be perfectly legitimate as part of a scientific research program. Consider this cherry picked list\footnote{There are nine specified in the original but I discuss only five: cherry picking!} of responses to a P-value being greater than 0.05 that have been described as P-hacking \citep{Motulsky:2014uv}: 

\begin{itemize}
	\item Analyze only a subset of the data; 
	\item Remove suspicious outliers; 
	\item Adjust data (e.g. divide by body weight); 
	\item Transform the data (i.e. logarithms); 
	\item Repeat to increase sample size (n); 
\end{itemize}

Before going any further I need to point out that Motulsky has a more realistic attitude to P-hacking than might be assumed from my treatment of his list. He writes: ``If you use any form of P-hacking, label the conclusions as `preliminary'.'' \citep[p. 1019]{Motulsky:2014uv}.

Analysis of only a subset of the data is illicit if the unanalysed portion is omitted in order to manipulate the P-value, but unproblematical if it is omitted for being irrelevant to the scientific question at hand. Removal of suspicious outliers is similar in being only sometimes inappropriate: it depends on what is meant by the term ``outlier''. If it indicates that a datum is a mistake such as a typographical or transcriptional error, then of course it should be removed (or corrected). If an outlier is the result of a technical failure of a particular run of the experimental then perhaps it should be removed, but the technical success or failure of an experimental run must not be judged by the influence of its data on the overall P-value. If with word outlier just denotes a datum that is further from the mean than the others in the dataset, then omit it at your peril! Omission of that type of outlier will reduce the variability in the data and give a lower P-value, but will markedly increase the risk of false positive results and it is, indeed, an illicit and damaging form of P-hacking.

Adjusting the data by standardisation is appropriate---desirable even---in some circumstances. For example, if a study concerns feeding or organ masses then standardising to body weight is probably a good idea. Such manipulation of data should be considered P-hacking only if an analyst finds a too large P-value in unstandardised data and then tries out various re-expressions of the data in search of a low P-value, and then reports the results as if that expression of the data was intended all along. The P-hackingness of log-transformation is similarly situationally dependent. Consider pharmacological $\textup{EC}_{50}$s or drug affinities: they are strictly bounded at zero and so their distributions are skewed. In fact the distributions are quite close to log-normal and so log-transformation before statistical analysis is appropriate and desirable. Log-transformation of $\textup{EC}_{50}$s gives more power to parametric tests and so it is common that significance testing of $\textup{logEC}_{50}$s gives lower P-values than significance testing of the un-transformed $\textup{EC}_{50}$s. An experienced analyst will choose the log-transformation because it is known from empirical and theoretical considerations that the transformation makes the data better match the expectations of a parametric statistical analysis. It might sensibly be categorised as P-hacking only if the log-transformation was selected with no justification other than it giving a low P-value.

The last form of P-hacking in the list requires a good deal more consideration than the others because, well, statistics is complicated. That consideration is facilitated by a concrete scenario---a scenario that might seem surprisingly realistic to some readers. Say you run an experiment with $n=5$ observations in each of two independent groups, one treated and one control, and obtain a P-value of 0.07 from Student's $t$-test. You might stop and integrate the very weak evidence against the null hypothesis into your inferential considerations, but you decide that more data will clarify the situation. Therefore you run some extra replicates of the experiment to obtain a total of $n=10$ observations in each group (including the initial 5), and find that the P-value for the data in aggregate is 0.002. The risk of the `significant' result being a false positive error is elevated because the data have had two chances to lead you to discard the null hypothesis. Conventional wisdom says that you have P-hacked. However, there is more to be considered before the experiment is discarded.

Conventional wisdom usually takes the global perspective. As mentioned above, it typically privileges false positive errors over any other consideration, and calls the procedure invalid. However, the extra data has added power to the experiment and lowered the expected P-value for any true effect size. From a local evidence point of view, increasing the sample increases the amount of evidence available for use in inference, which is a good thing. Is extending an experiment after the statistical analysis a good thing or a bad thing? The conventional answer is that it is a bad thing and so the conventional advice is don't do it! However, a better response might balance the bad effect of extending the experiment with the good. Consideration of the local and global aspects of statistical inference allows a much more nuanced answer. The procedure described would be perfectly acceptable for a preliminary experiment.

Technically the two-stage procedure in that scenario allows \textit{optional stopping}. The scenario is not explicit, but it can be discerned that the stopping rule was, in effect, run $n=5$ and inspect the P-value; if it is small enough then stop and make inferences about the null hypothesis; if the P-value is not small enough for the stop but nonetheless small enough to represent some evidence against the null hypothesis, add an extra 5 observations to each group to give $n=10$, stop, and analyse again. We do not know how low the interim P-value would have to be for the protocol to stop, and we do not know how high it could be and the extra data still be gathered, but no matter where those thresholds are set, such stopping rules yield false positive rates higher than the nominal critical value for stopping would suggest. Because of that, the conventional view (the global perspective, of course) is that the protocol is invalid, but it would be more accurate to say that such a protocol would be invalid unless the P-value or the threshold for a Neyman--Pearsonian dichotomous decision is adjusted as would be done with a formal \textit{sequential test}. It is interesting to note that the elevation of false positive rate is not necessarily large. Simulations of the scenario as specified and with P$<$0.1 as the threshold for continuing show that the overall false positive error rate would be about 0.008 when the the critical value for stopping at the first stage is 0.005, and about 0.06 when that critical value is 0.05.

The increased rate of false positives (global error rate) is real, but that doesn't mean that the evidential meaning of the final P-value of 0.002 is changed. It is the same local evidence against the null as if it was obtained from a simpler one stage protocol with $n=10$. After all, the data are \textit{exactly the same} as if the experimenter had intended to obtain $n=10$ from the beginning. The optional stopping has changed the global properties of the statistical procedure but not the local evidence which contained in the actualised data.

You might be wondering how it is possible that the local evidence be unaffected by a process that increases the global false positive error rate. The rationale is that the evidence is contained within the data but the error rate is a property of the procedure---evidence is local and error rates are global. Recall that false positive errors can only occur when the null hypothesis is true. If the null is true then the procedure has increased the risk of the data leading us to a false positive decision, but if the null is false then the procedure has \textit{decreased} the risk of a false negative decision. Which of those has paid out in this case cannot be known because we do not know the truth of this local null hypothesis. It might be argued that an increase in the global risk of false positive decisions should outweigh the decreased risk of false negatives, but that is a value judgement that ought to take into account particulars of the experiment in question, the role of that experiment in the overall study, and other contextual factors that are unspecified in the scenario and that vary from circumstance to circumstance. 

So, what can be said about the result of that scenario? The result of P=0.002 provides moderately strong evidence against the null hypothesis, but it was obtained from a procedure with sub-optimal false positive error characteristics. That sub-optimality should be accounted for in the inferences that made from the evidence, but it is only confusing to say that it alters the evidence itself, because it is the data that contain the evidence and the sub-optimality did not change the data. Motulsky provides good advice on what to do when your experiment has the optional stopping:

\begin{quote}
	\begin{itemize}
	\item For each figure or table, clearly state whether or not the sample size was chosen in advance, and whether every step used to process and analyze the data was planned as part of the experimental protocol.
	\item If you used any form of P-hacking, label the conclusions as ``preliminary.''	
	\end{itemize}
\end{quote}

Given that basic pharmacological experiments are often relatively inexpensive and quickly completed one can add to that list the option of also corroborating (or not) those results with a fresh experiment designed to have a larger sample size (remember the significance filter exaggeration machine) and performed according to the design. Once we move beyond the globalist mindset of one-and-done such an option will seem obvious. 

\subsection{What is a statistical model?}

I remind the reader that this chapter is written under the assumption that pharmacologists can be trusted to deal with the full complexity of statistics. That assumption gives me licence to discuss unfamiliar notions like the role of the statistical model in statistical analysis. All too often the statistical model is often invisible to ordinary users of statistics and that invisibility encourages thoughtless use of flawed and inappropriate models, thereby contributing to the misuse of inferential statistics like P-values.

A statistical model is what allows the formation of calibrated statistical inferences and non-trivial probabilistic statements in response to data. The model does that by assigning probabilities to potential arrangements of data. A statistical model can be thought of as a set of assumptions, although it might be more realistic to say that a chosen statistical model imposes a set of assumptions onto the experimenter.

\begin{quote}
I have often been struck by the extent to which most textbooks, on the flimsiest of evidence, will dismiss the substitution of assumptions for real knowledge as unimportant if it happens to be mathematically convenient to do so. Very few books seem to be frank about, or perhaps even aware of, how little the experimenter actually \textit{knows} about the distribution of errors in his observations, and about facts that are assumed to be known for the purposes of statistical calculations. \hfill---\citep[p. $v$]{ColquhounBook}	
\end{quote}

Statistical models can take a variety of forms \citep{McCullagh2002}, but the model for the familiar Student's $t$-test for independent samples is reasonably representative. That model consists of assumed distributions (normal) of two populations with parameters mean ($\mu_1$ and $\mu_2$) and standard deviation ($\sigma_1$ and $\sigma_2$),\footnote{The ordinary Student's $t$-test assumes that $\sigma_1=\sigma_2$, but the Welch-Scatterthwaite variant relaxes that assumption.} and a rule for obtaining samples (e.g. a randomly selected sample of $n=6$ observations from each population).  A specified value of the the difference between means serves as the null hypothesis, so $H_0: \mu_1-\mu_2=\delta_{H_0}$. The test statistic is\footnote{Oh no! An equation! Don't worry, it's the only one, and, anyway, it is too late now to stop reading.}
$$t=\frac{(\bar{x}_1-\bar{x}_2)-\delta_{H_0}}{s_p \sqrt{1/n_1+1/n_2}} $$ 
where $\bar{x}$ is a sample mean and $s_p$ is the pooled standard deviation. The explicit inclusion of a null hypothesis term in the equation for $t$ is relatively rare, but it is useful because it shows that the null hypothesis is just a possible value of the difference between means. Most commonly the null hypothesis says that the difference between means is zero---it can be called a `nill-null'---and in that case the omission of $\delta_{H_0}$ from the equation makes no numerical difference. 

Values of $t$ calculated by that equation have a known distribution when $\mu_1-\mu_2=\delta_{H_0}$, and that distribution is Student's $t$-distribution.\footnote{Technically it's the central Student's $t$-distribution. When $\delta \neq \delta_{H_0}$ it's a non-central $t$-distribution \citep{Cumming:2001eu}.} Because the distribution is known it is possible to define hypothesis test acceptance regions for any level of $\alpha$ for a hypothesis test, and any observed $t$-value can be converted into a P-value in a significance test.

An important problem that a pharmacologist is likely to face when using a statistical model is that it's just a model. Scientific inferences are usually intended to communicate something about the real world, not the mini world of a statistical model, and the connection between a model-based probability of obtaining a test statistic value and the state of the real world is always indirect and often inscrutable. Consider the meaning conveyed by an observed P-value of 0.002. It indicates that the data are strange or unusual compared to the expectations of the statistical model when the parameter of interest is set to the value specified by the null hypothesis. The statistical model expects a P-value of, say, 0.002 to occur only 2 times out of a thousand on average when the null is true. If such a P-value is observed then one of these situations has arisen:

\begin{itemize}
	\item a two in a thousand accident of random sampling has occurred;
	\item the null hypothesised parameter value is not close to the true value;
	\item the statistical model is flawed or inapplicable because one or more of the assumptions underlying its application are erroneous.
\end{itemize}

\begin{figure}
\begin{center}
\includegraphics[width=0.5\linewidth]{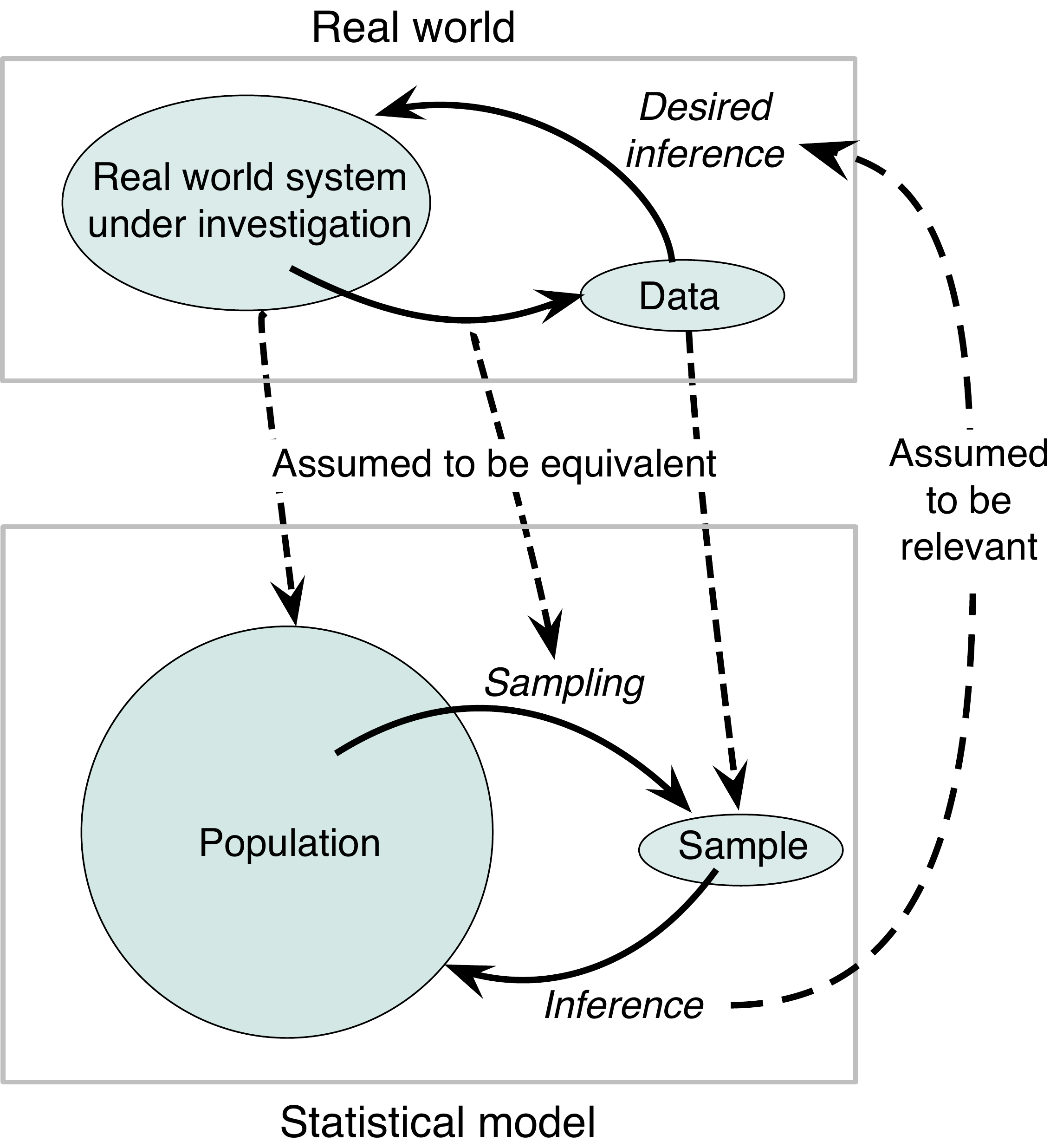}
\end{center}
\caption{
Diagram of inference using a statistical model.
}
\label{FigInfMed}
\end{figure}

Typically only the first and second are considered, but the last is every bit as important because when the statistical model is flawed or inapplicable then the expectations of the model are not relevant to the real world system that spawned the data. Figure \ref{FigInfMed} shows the issue diagrammatically. When we use that statistical inference to inform inferences about the real world we are implicitly assuming: (i) that the real world system that generated the data is an analog to the population in the statistical model; (ii) that the way the data were obtained is well described by the sampling rule of the statistical model; and (iii) that the observed data is analogous to the random sample assumed in the statistical model. To the degree that those assumptions are erroneous there is degradation of the relevance of the model-based statistical inference to the real world inference that is desired.

Considerations of model applicability are often limited to the population distribution (is my data normal enough to use a Student's $t$-test?) but it is much more important to consider whether there is a definable population that is relevant to the inferential objectives and whether the experimental units (``subjects'') approximate a random sample. Cell culture experiments are notorious for having ill-defined populations, and while experiments with animal tissues may have a definable population, the animals are typically delivered from an animal breeding or holding facility and are unlikely to be a random sample. Issues like those mean that the calibration of uncertainty offered by statistical methods might be more or less uncalibrated.  For good inferential performance in the real world, there has to be a flexible and well-considered linking of model-based statistical inferences and scientific inferences concerning the real world.

\section{P-values and inference}

A P-value tells you how well the data match with the expectations of a statistical model when the null hypothesis is true. But, as we have seen, there are many considerations that have to be made before a low P-value can safely be taken to provide sufficient reason to say that the null hypothesis is false. What's more, inferences about the null hypothesis are not always useful. Royall argues that there are three fundamental inferential questions that should be considered when making scientific inferences \citep{RoyallBook} (here paraphrased and re-ordered):

\begin{enumerate}
\item What do these data say?
\item What should I believe now that I have these data?
\item What should I do or decide now that I have these data?	
\end{enumerate}

Those questions are distinct, but not entirely independent and there is no single best way to answer to any of them.

A P-value from a significance test is an answer to the first question. It communicates how strongly the data argue against the null hypothesis, with a smaller P-value being a more insistent shout of ``I disagree!''. However, the answer provided by a P-value is at best incomplete, because it is tied to a particular null hypothesis within a particular statistical model and because it captures and communicates only some of the information that might be relevant to scientific inference. The limitations of a P-value can be thought of as analogous to a black and white photograph that captures the essence of a scene, but misses coloured detail that might be vital for a correct interpretation.

Likelihood functions provide more detail than P-values and so they can be superior to P-values as answers to the question of what the data say. However, they will be unfamiliar to most pharmacologists and they are not immune to problems relating to the relevance of the statistical model and the peculiarities of experimental protocol.\footnote{\cite{RoyallBook} and other proponents of likelihood-based inference (e.g. \cite{BergerWolpert}) make a contrary argument based on the likelihood principle and the (irrelevance of) sampling rule principle, but those arguments may fall down when viewed with the local versus global distinction in mind. Happily, those issues are beyond the scope of this chapter.} As this chapter is about P-values, we will not consider likelihoods any further, and those who, correctly, see that they might offer utility can read Royall's book \citep{RoyallBook}.

The second of Royall's questions, What should I believe now that I have these data?, requires integration of the evidence of the data with what was believed prior to the evidence being available. A formal statistical combination of the evidence with prior beliefs can be done using Bayesian methods, but they are rarely used for the analysis of basic pharmacological experiments and are outside the scope of this chapter about P-values. Considerations of belief can be assisted by P-values because when the data argue strongly against the null hypothesis one should be less inclined to believe it true, but it is important to realise that P-values do not in any way measure or communicate belief.

The Neyman--Pearsonian hypothesis test framework was devised specifically to answer the third question: it is a decision theoretic framework. Of course, it is a good decision procedure \textit{only} when $\alpha$ is specified prior to the data being available, and when a loss function informs the experimental design. And it is only useful when there is a singular decision to be made regarding a null hypothesis, as can be the case in acceptance sampling and in some randomised clinical trials. A singular decision regarding a null hypothesis is rarely a sufficient inference from the collection of experiments and observations that typically make up a basic pharmacological studies and so hypothesis tests should not be a default analytical tool (and the hybrid NHST should not be used in any circumstance).

Readers might feel that this section has failed to provide a clear method for making inferences about any of the three questions, and they would be correct. Statistics is a set of tools to help with inferences and not a set of inferential recipes, a scientific inferences concerning the real world have to be made by scientists, and my intention with this reckless guide to P-values is to encourage an approach to scientific inference that is more thoughtful than statistical significance. After all, those scientists invariably know much more than statistics does about the real world, and have a superior understanding of the system under study. Scientific inferences should be made after principled consideration of the available evidence, theory and, sometimes, informed opinion. A full evaluation of evidence will include both consideration of the strength of the local evidence and the global properties of the experimental system and statistical model from which that evidence was obtained. It is often difficult, just like statistics, and there is no recipe.

\bibliography{Lew_chapter.bib}

\end{document}